\documentclass[a4paper,12pt]{article}

\usepackage{amsmath}
\usepackage{amssymb}
\usepackage{amsthm}

\newcommand{\prt}{\partial}

\def\N{{\mathbb N}}
\def\Pr{{\mathbb P}}
\def\F{{\mathcal F}}

\def\ni{{\noindent}}

\def\be{\begin{equation}}
\def\ee{\end{equation}}

\newtheorem{theorem}{Theorem}

\theoremstyle{definition}

\begin{document}
\bibliographystyle{plain}

\title{{\Large\bf  Classical integrability for beta-ensembles and general Fokker-Planck equations.}}
\author{Igor Rumanov \\  
{\small Dept. of Applied Mathematics, CU Boulder, Boulder, CO} \\
{\small e-mail: igor.rumanov@colorado.edu} }

\maketitle

\bigskip

\begin{abstract}
Beta-ensembles of random matrices are naturally considered as quantum integrable systems, in particular, due to their relation with conformal field theory, and more recently appeared connection with quantized Painlev\'e Hamiltonians. Here we demonstrate that, at least for {\it even integer} beta, these systems are classically integrable, e.g.~there are Lax pairs associated with them, which we explicitly construct. To come to the result, we show that a solution of every Fokker-Planck equation in one space (and one time) dimensions can be considered as a component of an eigenvector of a Lax pair. The explicit finding of the Lax pair depends on finding a solution of a governing system -- a closed system of two nonlinear PDEs of hydrodynamic type. This result suggests that there must be a solution for all values of beta. We find the solution of this system for even integer beta in the particular case of quantum Painlev\'e II related to the soft edge of the spectrum for beta-ensembles. The solution is given in terms of Calogero system of $\beta/2$ particles in an additional time-dependent potential. Thus, we find another situation where quantum integrability is reduced to classical integrability. 
\end{abstract}

\newpage

\section{Introduction}

Beta-ensembles of random matrices were defined by Dyson~\cite{DyBeta} in terms of their eigenvalue probability distribution

$$
d\Pr(x) \sim \prod_{i<j}^n(x_i-x_j)^{\beta}\prod_{k=1}^ne^{-V(x_k)}dx_k,   \eqno(1.1)
$$

\ni which generalizes the eigenvalue distribution of three random matrix ensembles (RME) with rotation invariance of joint distribution of matrix entries: $\beta=1, 2, 4$ corresponds, respectively, to real symmetric, complex Hermitian and quaternionic self-dual (symplectic) RME. For the three ensembles, change of variables from matrix entries to eigenvalues-eigenvectors and integrating out the eigenvector part gives eq.~(1). Writing $|x_i-x_j|^{\beta} = \exp(\beta\log|x_i-x_j|)$, Dyson was first to interpret eq.~(1) for all values of $\beta \in (0, +\infty)$ as energy of gas of charged particles on the line with mutual Coulomb repulsion and external potential $V(x)$. This interpretation proved to be very fruitful and since then the model appeared in many physical situations, see e.g.~\cite{F2010} and references therein.
\par Its connection with Conformal Field Theory (CFT), important in gauge field and string theory as well as for quantitative description of two-dimensional critical phenomena (second order phase transitions for systems restricted to a plane), exhibits the prominent role of these ensembles. The Virasoro algebra, which governs the algebraic structure of CFT, has commutation relations

$$
[L_m, L_k] = (k-m)L_{m+k} + \frac{c}{12}(m^3-m)\delta_{m,-k},   \eqno(1.2)
$$

\ni where the original meaning of the generators $L_n$ is the Fourier components of stress-energy tensor in the corresponding CFT, and $c$ is its fundamental parameter called central charge, measuring effective vacuum energy density. As was noticed first likely by Awata et al.~\cite{AwMOSh}, integrals of eq.~(1.1) satisfy the Virasoro constraints, i.e.~the generators $V_k$, $k\ge -1$ of infinitesimal integration variable changes of the form $x_i \to x_i + \epsilon x_i^{k+1}$, $\epsilon \to 0$, satisfy relations (1.2) for $m, k \ge -1$, i.e.~satisfy a subalgebra of the Virasoro algebra. Although the central charge does not appear explicitly in this subalgebra, it is related to $\beta$ as~\cite{AwMOSh}

$$
c = 1 - 6\frac{(1-\beta/2)^2}{\beta/2},   \eqno(1.3)
$$

\ni which gives $c=1$ for $\beta=2$ and $c=-2$ for $\beta=1$ and $\beta=4$. The duality $\beta/2 \leftrightarrow 2/\beta$ is clear in eq.~(1.3) since these dual values of $\beta$ give the same central charge. The corresponding integrals for the three special values of $\beta$ are also known to be $\tau$-functions of classical integrable hierarchies like the KP or Toda lattice hierarchy. The last fact is not known to be true for the other values of $\beta$. The connection with CFT, however, implies that general beta-ensembles belong to the realm of quantum integrability. The multiple links between CFT and quantum integrable systems like quantum spin chains were found. Especially explicit were the constructions of Baxter $T$ and $Q$ operators immediately from CFT, given in~\cite{BLZ}. Most of these links and constructions are, however, complicated and non-rigorous so far. A recent appearance of integrable Benjamin-Ono hierarchy in connection with $\beta$-matrix models and associated Selberg-type integrals in the context of CFT~\cite{AFLT} is already an indication of hidden classical integrability. We are going to argue here that the recent developments in the theory of beta-ensembles lead to simpler and firmer such links. 
\par These more recent developments started with matrix realizations of Gaussian and Laguerre beta-ensembles discovered by Dumitriu and Edelman in 2002~\cite{DE02}. They found tridiagonal RME which give eq.~(1.1) as eigenvalue distribution with the corresponding Gaussian and Laguerre potentials $V(x)$ (see e.g.~\cite{ABG12, EdEtAlBeta13} for the latest developments in building full matrix models for beta-ensembles). Then Edelman and Sutton~\cite{EdSut} and, more rigorously, Ramirez, Rider and Virag~\cite{RRV} considered the large size $n$ limit of the tridiagonal RME with the soft edge of the spectrum scaling. In this limit, the random matrices were found to turn into a random Schr\"odinger Hamiltonian,

$$
H(t) = -\frac{d^2}{dt^2} + t + \frac{2}{\sqrt\beta}B'(t),    \eqno(1.4)
$$

\ni with $B'(t)$ being the white noise (i.e.~Airy Hamiltonian with the noise added). Rather than directly studying the eigenproblem $H\psi = \lambda\psi$ of (1.4), the authors of~\cite{RRV} introduced the logarithmic derivative $p(t)=\psi'(t)/\psi(t)$ as the new dependent function instead of $\psi$, and thereby obtained the first order (though nonlinear) stochastic PDE for $p$:

$$
dp = (t-\lambda-p^2)dt + \frac{2}{\sqrt\beta}dB(t),    \eqno(1.5)
$$

\ni This equation has the standard Langevin form, therefore it leads (see e.g.~\cite{Ok, RevYor}) to the corresponding diffusion-drift (or Fokker-Planck, or Kolmogorov) equation for the probability distribution $\F_{\beta}(t, x)$ of $p$, encoding all the statistical properties of eq.~(1.5):

$$
\left(\prt_t + \frac{2}{\beta}\prt_{xx} + (t-x^2)\prt_x\right)\F_{\beta}(t, x) = 0.   \eqno(1.6)
$$

\ni The boundary conditions ensure that the solution $\F_{\beta}$ to Fokker-Planck (FP) eq.~(1.6) is a probability distribution function:

$$
\F_{\beta}(t,x) \to 0  \ \ \ \text{as } x \to -\infty, t < \infty,  \ \ \ \ \   \F_{\beta}(t,x) \to 1 \ \ \ \text{as } t, x \to \infty \text{ together}.   \eqno(1.7)
$$

\ni This boundary value problem (BVP) has a unique bounded solution, which has also a limit $F_{\beta}(t) = \lim_{x\to\infty}\F_{\beta}(t, x)$. Functions $F_{\beta}(t)$ are the best available today definitions of $\beta$-Tracy-Widom (TW) distributions, the generalizations of the celebrated TW distributions~\cite{TW-Airy, TW-OrtSym} for $\beta=2, 1, 4$ derived from integrable theory. The above BVP has appeared for the first time explicitly in a slightly different context -- in consideration of {\it spiked} $\beta$-RME, see~\cite{BBP} for $\beta=2$ case, arising as sample covariance matrices in statistics, by Bloemendal and Virag~\cite{BV1}. There the $x$-variable has a somewhat different meaning: $\psi'(0) = x\psi(0)$ is the boundary condition for the random eigenfunctions of the stochastic Airy operator (1.4). The last authors explicitly established the connection of eq.~(1.6) with a Lax pair for the Painlev\'e II equation, albeit only for $\beta = 2$ and $4$. This particular Lax pair was found by Baik and Rains~\cite{BR01} from studying certain random growth models, its eigenfunctions are limits of orthogonal polynomials on the unit circle. The Lax pair for Painlev\'e II~\cite{BR01, BV1} is

$$
\prt_{t}\left(\begin{array}{c}f \\ g \end{array}\right) = \left(\begin{array}{cc}  0 & q \\ q & -x  \end{array}\right)\left(\begin{array}{c}f \\ g \end{array}\right),    \eqno(1.8) 
$$

$$
\prt_{x}\left(\begin{array}{c}f \\ g \end{array}\right) = \left(\begin{array}{cc}  q^2 & -qx-q' \\ -qx+q' & x^2 - t - q^2  \end{array}\right)\left(\begin{array}{c}f \\ g \end{array}\right),    \eqno(1.9)
$$

\ni where $q(t)$ is the Hastings-McLeod solution of Painlev\'e II~\cite{HMcL80, TW-Airy}. A still mysterious fact is that no such simple connection seem to exist for the third special value of $\beta$, $\beta=1$. Moreover, there is another large $n$ limit for $\beta$-ensembles, the so-called hard-edge limit. The limiting Schr\"odinger operator as well as the corresponding FP equation have been found for this case by Ramirez and Rider in~\cite{RR08}, see also~\cite{RRZ}. Here again the connection with Lax pairs, this time for Painlev\'e III equation, was found only for $\beta=2, 4$ but not for $\beta=1$~\cite{HBP3}.
\par Equation (1.6) has a clear meaning in quantum integrable theory. It is the imaginary time Schr\"odinger equation with canonically quantized Painlev\'e II Hamiltonian. This way, identification of $2/\beta$ with the Planck constant $\hbar$ can be made. The same is true of the hard-edge FP operator of~\cite{RR08} corresponding to quantum Painlev\'e III Hamiltonian. Recently the non-stationary Schr\"odinger (aka Fokker-Planck) equations for all six Painlev\'e equations have been considered by Nagoya~\cite{Nag11, Nag11H} who found their particular solutions in the form of matrix integrals for beta-ensembles with certain special potentials. A remarkable fact is that for those ensembles $\hbar=\beta/2$ unlike in our case, an explanation of this is given in the review~\cite{BetaGarnier}.
\par Classical integrability for general beta-ensembles and related models remained elusive, see e.g.~a recent discussion of these issues in the context of gauge field and string theory~\cite{MMShRes}. Here we find an exact quantum-classical correspondence for {\it all even} $\beta$. 
\par A source of inspiration for the current work comes from the results of Krichever et al.~\cite{KLWZ97} which have a number of recent developments, see e.g.~\cite{ZaBetheHir} and references therein. There the authors found that certain quantum transfer matrix eigenvalues were classical $\tau$-functions since they satisfied the discrete bilinear Hirota equation~\cite{Hir81}.
\par We are going to make it plausible that an exact quantum-classical correspondence exists for all $\beta$. The probability distribution $\F_{\beta}(t,x)$ can be represented as a component of a $2\times2$ Lax pair eigenvector for all $\beta$. Below we explicitly find such a Lax pair for the case of soft edge for $\beta$-ensembles (i.e.~for quantum Painlev\'e II) when $\beta$ is an even integer. In fact, we show more: {\it any} Fokker-Planck (FP) equation in one space dimension can be made such an eigenvector component. Requiring this leads to a closed system of two nonlinear PDEs of hydrodynamic type with viscosity, which we call governing equations. As we will see, their solution may turn out to be simpler than the solution of the FP equation one started with. This gives additional justification for the approach taken here.
\par A similar approach was put forward recently by Zabrodin and Zotov~\cite{ZZFP}. One difference is that these authors restrict their Lax pairs to be traceless, which we do not do. More importantly, however, we stress that the approach can be applied e.g.~to $\beta$-ensembles for all $\beta$ (this is not clear from~\cite{ZZFP}). In fact, the solutions these authors find, correspond to the classical Painlev\'e equations, i.e.~to the special values of $\beta$: $\beta=2$ (the one-pole solutions) and $\beta=4$ (the two-pole solutions), as can be seen from the results of~\cite{BV1, HBP3} for Painlev\'e II and III. Nevertheless, considerations of~\cite{ZZFP} helped us come up with our ansatz for even $\beta$, see section ``Quantum Painlev\'e II" below. 
\par The history here is, however, far longer. An approach, in some respects even more similar to ours, appeared in the paper by Bluman and Cole~\cite{BC69} back in 1969, where the particular case of our governing system, corresponding to the simple heat equation in place of general FP, was written down. See e.g.~\cite{MansHeat99, ArEtAl2010} for later developments. Here we demonstrate a wider applicability of the approach started in~\cite{BC69}, its relation with integrable theory, see also~\cite{ZenSan08} on the last topic.
\par Besides, a function simply related to $\F_2$ of eq.~(1.6) appeared in~\cite{NaKaMaP2} where log-derivative of the first eigenvector component $Y_1$ of traceless Lax pair for Painlev\'e II~\cite{FN80, JM81} was introduced as a generating function of entries of Hankel determinant {\it generic} solutions of Painlev\'e II and its connection with Toda chain was exposed ($\F_2=F_2(t)\exp(-1/2(tx-x^3/3))Y_1$ in our normalization). 
\par Plan of the paper is as follows. In section 2 we derive the connection of a general FP equation with Lax pairs, which reduces to the necessity of solving certain closed coupled system of two nonlinear PDEs -- the governing system. In section 3 the solution of the governing system for quantum Painlev\'e II is obtained for even integer $\beta$. An example Lax pair for even $\beta$ is constructed in section 4 according to the general formulas of subsection 2.1. The last section 5 contains a brief discussion of problems for future work. Some technical details are presented in the appendices.

\section{FP equations and Lax pairs}

Consider a general Fokker-Planck (FP) equation (an example to keep in mind is eq.~(1.6) with rescaled $x$ and $t$ so that $\kappa=\beta/2$):

$$
(\kappa\prt_t + \sigma(t,x)\prt_{xx} + v(t,x)\prt_x + \alpha(t,x))\F_{\kappa}(t,x) = 0.   \eqno(2.1)
$$

\ni We want its solution to be the (say, first) component of the eigenvector of a $2\times 2$ Lax pair,

$$
\prt_x\left(\begin{array}{c}\F_{\kappa} \\ G \end{array}\right) = L\left(\begin{array}{c}\F_{\kappa} \\ G \end{array}\right),  \ \ \ \ \ \   \prt_t\left(\begin{array}{c}\F_{\kappa} \\ G \end{array}\right) = B\left(\begin{array}{c}\F_{\kappa} \\ G \end{array}\right),  \eqno(2.2)   
$$

\ni where we denote

$$
L = \left(\begin{array}{cc} L_1 & L_+ \\ L_- & L_2 \end{array}\right),   \ \ \ \ \ B = \left(\begin{array}{cc} B_1 & B_+ \\ B_- & B_2 \end{array}\right).
$$

\ni Then the first components of these equations are

$$
\prt_t\F_{\kappa} = B_1\F_{\kappa} + B_+G, \ \ \ \ \ \prt_x\F_{\kappa} = L_1\F_{\kappa} + L_+G,   \eqno(2.3)
$$

\ni and, eliminating $G$ from them, one obtains another, first-order, PDE for $\F_{\kappa}$:

$$
\prt_t\F_{\kappa} - b_+\prt_x\F_{\kappa} + b_1\F_{\kappa} = 0,   \eqno(2.4)
$$

\ni where we denoted

$$
b_+ = \frac{B_+}{L_+}, \ \ \ \ \ b_1 = b_+L_1-B_1.   \eqno(2.5)
$$

\ni Eliminating $\prt_t\F_{\kappa}$ from (2.1) and (2.4), one sees that $\F_{\kappa}$ satisfies also an ODE in $x$:

$$
(\sigma\prt_{xx} + (v+\kappa b_+)\prt_x + (\alpha-\kappa b_1))\F_{\kappa} = 0.  \eqno(2.6)
$$

\ni Thus, finding a Lax pair (2.2) amounts to effective separation of variables and getting ODEs for $\F_{\kappa}$. An example when all this is known to be true is the function $f$ of Baik-Rains~\cite{BR01} Lax pair (but there $\kappa=1$), see~\cite{BaikP06} about properties of $f$. 
\par We next express $\prt_{xx}\F_{\kappa}$, using (2.2), as

$$
\prt_{xx}\F_{\kappa} = (\prt_xL_1 + L_1^2 + L_+L_-)\F_{\kappa} + (\prt_xL_+ + L_tL_+)G,   \eqno(2.7)
$$

\ni where we denoted $L_t = L_1+L_2$. Then FP eq.~(2.1) together with (2.3) and (2.7) leads to the two nontrivial constraints for the components of the Lax matrices:

$$
\kappa B_1 + \sigma(\prt_xL_1 + L_1^2 + L_+L_-) + vL_1 + \alpha = 0,   \eqno(2.8)  
$$

$$
\kappa B_+ + \sigma(\prt_xL_+ + L_tL_+) + vL_+ = 0.   \eqno(2.9)  
$$

\ni Besides, we have the zero curvature (Lax) equations:

$$
\prt_tL_1 = \prt_xB_1 + B_+L_- - B_-L_+,  \eqno(2.10)
$$

$$
\prt_tL_+ = \prt_xB_+ + B_dL_+ - B_+L_d,  \eqno(2.11)
$$

$$
\prt_tL_- = \prt_xB_- + B_-L_d - B_dL_-,  \eqno(2.12)
$$

$$
\prt_tL_t = \prt_xB_t,            \eqno(2.13)             
$$

\ni where we also introduced $L_d=L_1-L_2$, $B_d=B_1-B_2$ and $B_t = B_1+B_2$. Therefore we have six independent equations for eight independent components of the matrices $L$ and $B$. Two of the components therefore remain arbitrary (possible to choose at will), which corresponds to the non-uniqueness of the second eigenvector component $G$. 
\par To transform the last six equations to a convenient form, we introduce further definitions:

$$
l_- = L_+L_-, \ \ \ \ \ b_- =  B_-L_+ - B_+L_- = B_-L_+ - b_+l_- , \ \ \ \ \ X_t = L_t + \frac{\prt_xL_+}{L_+},    
$$

$$
y = \frac{\prt_tL_+}{L_+} - B_d + 2B_1 = \frac{\prt_tL_+}{L_+} + B_t,  \ \ \ \ \ X_1 = l_- + L_1^2 + \prt_xL_1 - X_tL_1.   \eqno(2.14)
$$

\ni Then eqs.~(2.8)--(2.13) are equivalent to the following system:

$$
-\kappa b_1 + \sigma X_1 + \alpha = 0,   \eqno(2.15)                     
$$

$$
\kappa b_+ + \sigma X_t + v = 0.   \eqno(2.16)  
$$

$$
\prt_tL_1 = \prt_xB_1 - b_-,  \eqno(2.17)
$$

$$
y = \prt_xb_+ - b_+(2L_1-X_t) + 2B_1 = \prt_xb_+ + b_+X_t - 2b_1,  \eqno(2.18)
$$

$$
\prt_tl_- = \prt_xb_- + (2L_1-X_t)b_- + b_+\prt_xl_- + 2\prt_xb_+\cdot l_-,  \eqno(2.19)
$$

$$
\prt_tX_t = \prt_xy.  \eqno(2.20)
$$

\ni Differentiating eq.~(2.16) w.r.t.~$t$ and using eqs.~(2.20), (2.18) and (2.16) to eliminate $\prt_tX_t$, $y$ and $X_t$, we obtain the first equation involving the functions $b_+$ and $b_1$ only:

$$
\kappa\prt_tb_+ + \sigma\prt_{xx}b_+ - \prt_x(b_+(\kappa b_+ + v)) + \left(\frac{\prt_x\sigma}{\sigma}b_+ - \frac{\prt_t\sigma}{\sigma}\right)(\kappa b_+ + v) + \prt_tv - 2\sigma\prt_xb_1 = 0.  \eqno(2.21)  
$$

\ni Differentiating $X_1$ w.r.t.~$t$, using its expression from (2.14) and substituting $t$-derivatives from eqs.~(2.19), (2.17) and (2.20), as well as expressions for $y$ from eq.~(2.18) and for $B_1$ from the second formula of (2.5), we obtain

$$
\prt_tX_1 = b_+\prt_xX_1 + 2\prt_xb_+\cdot X_1 - \prt_{xx}b_1 + X_t\prt_xb_1.   \eqno(2.22)
$$

\ni Then, using (2.15) and (2.16) to eliminate, respectively, $X_1$ and $X_t$ from (2.22), we obtain the second equation involving only the governing functions $b_1$ and $b_+$:


$$
\prt_t\left(\frac{\kappa b_1 - \alpha}{\sigma}\right) + \prt_{xx}b_1 - b_+\prt_x\left(\frac{\kappa b_1 - \alpha}{\sigma}\right) + \left(\frac{\kappa b_+ + v}{\sigma}\right)\prt_xb_1 = 2\left(\frac{\kappa b_1 - \alpha}{\sigma}\right)\prt_xb_+.   \eqno(2.23)  
$$

\ni Eq.~(2.21) can be written as a conservation law:   

$$
\prt_t\left(\frac{\kappa b_+ + v}{\sigma}\right) + \prt_x\left(\prt_xb_+ - \frac{b_+(\kappa b_+ + v)}{\sigma} - 2b_1\right) = 0   \eqno(2.24)  
$$

\ni (it comes from the trace equation (2.13) or (2.20)). Eq.~(2.23) is not a conservation law in general. It can be also written in another form where only the nonlinear terms remaining after a cancellation are present: 


$$
\kappa\prt_tb_1 + \sigma\prt_{xx}b_1 + v\prt_xb_1 = \left(2\prt_xb_+ + \frac{\prt_t\sigma}{\sigma} - \frac{\prt_x\sigma}{\sigma}b_+\right)(\kappa b_1 - \alpha) + \prt_t\alpha - b_+\prt_x\alpha.   \eqno(2.25)
$$

\ni Thus, we obtain the first main result:

\begin{theorem}
Equations (2.24) and (2.23) (or (2.24 and (2.25)) constitute a closed system of two nonlinear PDEs, solvability of which ensures the relation of the general FP equation (2.1) and its solutions with classical Lax pairs and their eigenfunctions. The other components of the Lax matrices $L$ and $B$ can be found from $b_+$ and $b_1$ algebraically, using the relations (2.5) and (2.14), up to the freedom of choosing two of them arbitrarily. 
\end{theorem}

\subsection{Explicit Lax Pairs}

If a solution of the governing equations (2.24), (2.25) for $b_+$ and $b_1$ is found, then the full Lax pair can be restored, with the remaining freedom to choose two arbitrary functions of the entries of $L$ and $B$. This restoration is particularly simple if one chooses the entries $L_+$ and either $L_1$ or $B_1$ as arbitrary. Then the other components are determined by algebraic formulas below:

$$
B_+ = b_+L_+,   \eqno(2.26)   
$$

$$
L_t \equiv L_1 + L_2 = -\frac{\kappa b_+ + v}{\sigma} - \frac{\prt_xL_+}{L_+},   \eqno(2.27)  
$$

$$
B_t \equiv B_1 + B_2 = \prt_xb_+ - \frac{b_+(\kappa b_+ + v)}{\sigma} - 2b_1 - \frac{\prt_tL_+}{L_+},   \eqno(2.28)  
$$

$$
b_+L_1 - B_1 = b_1,   \eqno(2.29)   
$$

$$
L_- = \frac{1}{L_+}\left(\frac{\kappa b_1 - \alpha - (\kappa b_+ + v)L_1}{\sigma} - L_1^2 - \prt_xL_1\right),   \eqno(2.30)   
$$

$$
B_- = b_+L_- + \frac{\prt_xB_1 - \prt_tL_1}{L_+}.   \eqno(2.31)   
$$

\ni The above formulas follow from eqs.~(2.16)--(2.18), (2.24) and definitions (2.5), (2.14). Let us note that the work~\cite{ZZFP} uses a different choice, which makes the Lax matrices traceless, i.e.~$B_t = L_t = 0$ in our notation. In this case $L_+$ is not arbitrary anymore (it is determined by eqs.~(2.27), (2.28) and the governing equations (2.24), (2.25)) and only one function (either $L_1$ or $B_1$) remains arbitrary, at the expense of making the Lax pairs not the most general possible. This ``choice of gauge" is not convenient for our purposes since e.g.~for beta-ensemble spectral probabilities there are known Lax pairs~\cite{BV1, HBP3} for special values of $\beta$, $\beta=2, 4$, not satisfying the traceless property.




\section{Quantum Painlev\'e II}

The Quantum Painlev\'e II equation appeared in the study of $\beta$-ensembles in~\cite{BV1} and later was identified as nonstationary (imaginary time) Schr\"odinger equation with canonically quantized Painlev\'e II Hamiltonian in~\cite{Nag11}. It corresponds to the case $\sigma\equiv 1$, $v = t-x^2$, $\alpha\equiv 0$ in eq.~(2.1), which then becomes 

$$
\left(\kappa\prt_t + \prt_{xx} + (t-x^2)\prt_x\right)\F(t, x) = 0,   \eqno(3.0)
$$

\ni i.e.~eq.~(1.6) with $t$ and $x$ rescaled as $x\to x/\kappa^{1/3}$, $t\to t/\kappa^{2/3}$, $\kappa=\beta/2$. Thus, the governing equations take the following form:

$$
\prt_t\left(\kappa b_+ + v\right) + \prt_x\left(\prt_xb_+ - b_+(\kappa b_+ + v) - 2b_1\right) = 0,   \eqno(3.1)  
$$

$$
\prt_t\left(\kappa b_1\right) + \prt_{xx}b_1 + v\prt_xb_1 = 2\kappa b_1\prt_xb_+.   \eqno(3.2)  
$$

\ni Considering asymptotics of the system as $x\to\infty$ one can see that, for all $\kappa$, eqs.~(3.1), (3.2) are compatible with

$$
b_+ \sim -\frac{1}{x} + o(1/x).    \eqno(3.3)  
$$

\ni On the other hand, at finite $x$, one can verify that, for all $\kappa$, eqs.~(3.1) and (3.2) admit Laurent expansions of $b_+$ and $b_1$ in $x$ which are either regular or have the form

$$
b_+ = -\frac{1}{\kappa}\frac{1}{x-Q(t)} + \sum_{k=0}^{\infty}Q_{+k}(t)(x-Q(t))^k,   \ \ \ \ \ b_1 = \frac{Q_1(t)}{x-Q(t)} + \sum_{k=0}^{\infty}Q_{1k}(t)(x-Q(t))^k,   \eqno(3.4)  
$$

\ni with a finite pole $Q(t)$ remaining undetermined. Now recall that the Lax pairs are known explicitly for the special cases $\kappa=1$ and $2$~\cite{BV1, HBP3}. For these cases, $b_+$ and $b_1$ are 

$$
\kappa=1: \ \ \ b_+ = -\frac{1}{x + q'(t)/q}, \ \ \ \ \ b_1 = -\frac{q^2}{x + q'(t)/q} - u(t),   \eqno(3.5)  
$$

$$
\kappa=2: \ \ b_+ = -\frac{x}{x^2-t-2q'-2q^2} = -\frac{1}{2(x-\sqrt{2q'+2q^2+t})}-\frac{1}{2(x+\sqrt{2q'+2q^2+t})}, 
$$

$$
b_1 = -\frac{(q'+q)x - q(2q'+2q^2+t)}{x^2-t-2q'-2q^2} + \frac{q-u}{2},   \eqno(3.6)  
$$

\ni where $q(t)$ is the Hastings-McLeod solution of Painlev\'e II and $u(t)$ is such that $u'(t) = -q^2$. All the above facts suggest to try the ansatz of the form

$$
b_+(t,x) = -\frac{1}{\kappa}\sum_{k=1}^N\frac{1}{x-Q_k(t)}.   \eqno(3.7)  
$$

\ni It is consistent with eq.~(3.3) only for $N=\kappa$, which implies that the ansatz can be true only for integer $\kappa$, i.e.~for even integer $\beta$. One has many indications from random matrix theory that the even integer values of $\beta$ are special, see e.g.~\cite{ForEvenBeta09}. As we will see shortly, the ansatz works exactly when $N=\kappa$.
\par From eq.~(3.7), one obtains

$$
\prt_t\rho \equiv \prt_t(\kappa b_+ + v) = -\sum_{k=1}^N\frac{Q_k'(t)}{(x-Q_k)^2} + 1,   \eqno(3.8)  
$$

$$
\prt_xb_+ = \frac{1}{\kappa}\sum_{k=1}^N\frac{1}{(x-Q_k)^2}.  \eqno(3.9)  
$$

\ni Let

$$
J(t,x) = \prt_xb_+ - b_+\rho - 2b_1,   \eqno(3.10)  
$$

\ni then, substituting eq.~(3.8) into eq.~(3.1), one finds the general solution for $J$,

$$
J(t,x) = -\sum_{k=1}^N\frac{Q_k'}{x-Q_k} - \frac{1}{N}\sum_{k=1}^N(x-Q_k) + J_0(t),   \eqno(3.11)  
$$

\ni where function $J_0(t)$ is arbitrary so far. From eqs.~(3.10), (3.11) one obtains

$$
2b_1(t, x) = \prt_xb_+ - b_+\rho - J = -\frac{1}{\kappa}\sum_{k=1}^N\sum_{j\neq k}^N\frac{1}{(x-Q_k)(x-Q_j)} + \frac{1}{\kappa}\sum_{k=1}^N\frac{\kappa Q_k' + t - Q_k^2}{x-Q_k} - 
$$

$$
- \frac{2}{\kappa}\sum_{k=1}^NQ_k - J_0(t) + \left(\frac{1}{N} - \frac{1}{\kappa}\right)\sum_{k=1}^N(x-Q_k),   \eqno(3.12)  
$$

\ni where we used that $v=t-x^2$ can be written as

$$
v(t,x) = t - Q_k^2 - 2Q_k(x-Q_k) - (x-Q_k)^2.   \eqno(3.13)  
$$

\ni Eq.~(3.12) can be rewritten as

$$
 2b_1(t, x) = \frac{1}{\kappa}\sum_{k=1}^N\frac{\kappa Q_k' + t - Q_k^2 - 2R_k}{x-Q_k} - \frac{2}{\kappa}\sum_{k=1}^NQ_k - J_0(t) + \left(\frac{1}{N} - \frac{1}{\kappa}\right)\sum_{k=1}^N(x-Q_k),   \eqno(3.14)  
$$

\ni where we denoted

$$
R_k = \sum_{j\neq k}^N\frac{1}{Q_k - Q_j}.   \eqno(3.15)  
$$

\ni Now we evaluate eq.~(3.2). Using eqs.~(3.14), (3.13), and (3.9), we find 

$$
2\prt_xb_1 = -\frac{1}{\kappa}\sum_{k=1}^N\frac{A_k}{(x-Q_k)^2} + \left(1 - \frac{N}{\kappa}\right),   \eqno(3.16)  
$$

$$
2\prt_{xx}b_1 = \frac{2}{\kappa}\sum_{k=1}^N\frac{A_k}{(x-Q_k)^3},   \eqno(3.17)  
$$

$$
2\kappa\prt_tb_1(t, x) = \sum_{k=1}^N\frac{Q_k'A_k}{(x-Q_k)^2} + \sum_{k=1}^N\frac{A_k'}{x-Q_k} - 
$$

$$
- 2\sum_{k=1}^NQ_k' - \kappa J_0'(t) - \left(\frac{\kappa}{N} - 1\right)\sum_{k=1}^NQ_k',   \eqno(3.18)  
$$

$$
2v\prt_xb_1 = -\frac{1}{\kappa}\sum_{k=1}^N\frac{(t-Q_k^2)A_k}{(x-Q_k)^2} + \frac{2}{\kappa}\sum_{k=1}^N\frac{Q_kA_k}{x-Q_k} + 
$$

$$
+ \frac{1}{\kappa}\sum_{k=1}^NA_k + \left(1 - \frac{N}{\kappa}\right)(t-Q_k^2) - 
$$

$$
- \left(\frac{1}{N} - \frac{1}{\kappa}\right)\sum_{k=1}^N2Q_k(x-Q_k) - \left(\frac{1}{N} - \frac{1}{\kappa}\right)\sum_{k=1}^N(x-Q_k)^2,   \eqno(3.19)  
$$

$$
2\kappa \prt_xb_+\cdot 2b_1 = \frac{2}{\kappa}\sum_{k=1}^N\frac{A_k}{(x-Q_k)^3} + 
$$

$$
+ \frac{2}{\kappa}\sum_{k=1}^N\frac{1}{(x-Q_k)^2}\left(\sum_{j\neq k}^N\frac{A_j}{Q_k-Q_j} - 2\sum_{k=1}^NQ_k - \kappa J_0(t) + \left(\frac{1}{N} - \frac{1}{\kappa}\right)(Q_k-Q_j)\right) +
$$ 

$$
+ \frac{2}{\kappa}\sum_{k=1}^N\frac{1}{x-Q_k}\left(\sum_{j\neq k}^N\frac{A_k-A_j}{(Q_k-Q_j)^2} + (\kappa-N)\right),   \eqno(3.20)  
$$

\ni where we introduced

$$
A_k = \kappa Q_k' + t - Q_k^2 - 2R_k.   \eqno(3.21)  
$$

\ni Thus, we have terms of powers $-3$ through $2$ in $(x-Q_k)$ for each $k$. Powers $-3$ cancel identically, while canceling powers $1$ and $2$ requires $N=\kappa$, as we anticipated, and we set $N=\kappa$ from now on. The other three powers give then nontrivial equations:

$$
(x-Q_k)^{-2}: \ \ \ \ \ (\kappa Q_k' - t + Q_k^2)A_k = 2\left(\sum_{j\neq k}^{\kappa}\frac{A_j}{Q_k-Q_j} - 2\sum_{k=1}^{\kappa}Q_k - \kappa J_0(t)\right),   \eqno(3.22)  
$$

$$
(x-Q_k)^{-1}: \ \ \ \ \ \kappa A_k' = -2Q_kA_k + 2\sum_{j\neq k}^{\kappa}\frac{A_k-A_j}{(Q_k-Q_j)^2},   \eqno(3.23)  
$$

$$
(x-Q_k)^0: \ \ \ \ \ \kappa J_0' + 2\sum_{k=1}^{\kappa}Q_k' = \frac{1}{\kappa}\sum_{k=1}^{\kappa}A_k.   \eqno(3.24)  
$$

\ni We apparently obtain $2\kappa + 1$ equations for $\kappa+1$ variables $Q_k, k=1, \dots, \kappa$ and $J_0$. They, however, are in fact not independent: eqs.~(3.23) follow from eqs.~(3.22) and (3.21) by differentiating eqs.~(3.22) w.r.t. $t$. This is proved in Appendix A.
\par Recalling eq.~(3.21) and using the fact that

$$
\sum_{k=1}^{\kappa}R_k = \sum_{k=1}^{\kappa}\sum_{j\neq k}^{\kappa}\frac{1}{Q_k - Q_j} = 0,   \eqno(3.25)   
$$

\ni one can rewrite eq.~(3.24) as

$$
\kappa J_0' + \sum_{k=1}^{\kappa}Q_k' = t - \frac{1}{\kappa}\sum_{k=1}^{\kappa}Q_k^2.  \eqno(3.26)   
$$

\ni If one introduces function $U(t)$ such that

$$
\kappa U'(t) = -\sum_{k=1}^{\kappa}Q_k^2,   \eqno(3.27)    
$$

\ni then eq.~(3.26) ``integrates" to

$$
\kappa J_0 + \sum_{k=1}^{\kappa}Q_k = \frac{t^2}{2} + U(t),    \eqno(3.28)   
$$

\ni (i.e.~eq.~(3.24) is equivalent to eqs.~(3.27), (3.28)). Then eqs.~(3.23) and (3.22) are expressed explicitly in terms of $\{Q_k, k=1,\dots,\kappa\}$ as follows, the details are in Appendix B. Eq.~(3.23) in fact represents the equations of motion for $\kappa$ particles with coordinates $Q_k, k=1, \dots, \kappa$:

$$
\kappa^2 Q_k'' = -2Q_k(t-Q_k^2) + \kappa-2 - \sum_{j\neq k}\frac{8}{(Q_k-Q_j)^3},  \eqno(3.29)   
$$

\ni and eq.~(3.22) gives first integrals of equations (3.29) together with eq.~(3.27):

$$
\frac{(\kappa Q_k')^2}{2} + tQ_k^2 - \frac{Q_k^4}{2} - (\kappa-2)Q_k - \sum_{j\neq k}^{\kappa}\frac{2}{(Q_k - Q_j)^2} + U(t) - 
$$

$$
- \sum_{j\neq k}^{\kappa}\frac{\kappa Q_k' + \kappa Q_j'}{Q_k - Q_j} + \sum_{j\neq k}^{\kappa}\sum_{l\neq k,j}^{\kappa}\frac{2}{(Q_k - Q_j)(Q_j - Q_l)} = 0.  \eqno(3.30)   
$$


\ni Eqs.~(3.30), (3.29) and (3.27) are the (second) {\bf main result} of the paper. If we sum eqs.~(3.30) over all $k$, then the sums of their last two terms are both zero, and we obtain

$$
\frac{1}{\kappa}\sum_{k=1}^{\kappa}\left(\frac{(\kappa Q_k')^2}{2} + tQ_k^2 - \frac{Q_k^4}{2} - (\kappa-2)Q_k - \sum_{j\neq k}^{\kappa}\frac{2}{(Q_k-Q_j)^2}\right) + U(t) = 0,  \eqno(3.31)  
$$

\ni which can be seen as a higher analog of the well-known relation for a Painlev\'e II, $q''(t)=tq+2q^3$, solution $q$ and function $u$ such that $u'(t) = -q^2$: $u = (q')^2-tq^2-q^4$.
\par To summarize, a Calogero system of $\kappa=\beta/2$ particles with coordinates $Q_k$, $k=1,\dots,\kappa$, in an additional external time-dependent potential appears here. Its equations of motion are eqs.~(3.29) and their first integrals are given by the equalities of the left-hand sides of eqs.~(3.30). Thus, there are $\kappa-1$ independent first integrals for the equations of motion (3.29). A complete description of the system is, however, given by $\kappa+1$ functions, $\{Q_k, k=1, \dots, \kappa\}$ and $U$. Their equations of motion are eqs.~(3.29) and (3.27), which have $\kappa$ independent first integrals eqs.~(3.30), or their $\kappa-1$ independent combinations involving $Q_k$-s only plus the equation (3.31). Function $U(t)$ can be also interpreted as time dependent Hamiltonian for the system of $\kappa$ interacting particles with coordinates $Q_k$ in external time-dependent field. 
\par {\it Remark.} One can rewrite eqs.~(3.30) more compactly if one introduces the {\it exchange operators} $\hat E_{kj} = \hat E_{jk}$ acting on $Q$-variables and their functions by exchange of $Q_k$ and $Q_j$ and leaving the other coordinates intact:

$$
\hat E_{kj} f(Q_1, \dots, Q_j, \dots, Q_k, \dots, Q_{\kappa}) = f(Q_1, \dots, Q_k, \dots, Q_j, \dots, Q_{\kappa}) \hat E_{kj}.
$$

\ni Then, if one introduces generalized momentum operators

$$
\hat P_k = \kappa Q_k' - \sum_{j\neq k}^{\kappa}\frac{2}{Q_k - Q_j} \hat E_{kj},    
$$

\ni one can verify that eqs.~(3.30) read simply

$$
\frac{\hat P_k^2}{2} + tQ_k^2 - \frac{Q_k^4}{2} - (\kappa-2)Q_k + U = 0,
$$

\ni and the operators $\hat P_k$ commute: $[\hat P_k, \hat P_j] = 0$ for any $j, k$.


\section{Explicit Lax pair for quantum Painlev\'e II, even $\beta$}

According to the formulas of subsection 2.1, one has the freedom to choose two arbitrary functions of $x$ and $t$ among the entries of matrices $L$ and $B$. It turns out that the choice for these functions described below is the most convenient and in a sense most natural for the case of integer values of $\kappa$ considered here. One can make the entry $L_+$ one of the functions to choose and specify it as 

$$
L_+ = \phi(t)\prod_{k=1}^{\kappa}(x-Q_k(t)),   \eqno(4.1)    
$$

\ni to match with the known pairs for the simplest cases $\kappa=1, 2$~\cite{HBP3}; $\phi(t)$ remains arbitrary. Then formulas (2.26) and (2.27) of section 2.1 determine, respectively, 

$$
B_+ = b_+L_+ = -\frac{\phi(t)}{\kappa}\sum_{k=1}^{\kappa}\prod_{j\neq k}^{\kappa}(x-Q_j(t)) = -\frac{\prt_xL_+}{\kappa},   \eqno(4.2)    
$$

$$
L_t \equiv L_1 + L_2 = -(\kappa b_+ + v) - \frac{\prt_xL_+}{L_+} = x^2 - t = -v.   \eqno(4.3)   
$$

\ni Formula (2.28) together with the expression (3.11) for $J(t,x)$ yield

$$
B_t \equiv B_1 + B_2 = \prt_xb_+ - b_+(\kappa b_+ + v) - 2b_1 - \frac{\prt_tL_+}{L_+} = J(t,x) - \frac{\prt_tL_+}{L_+} = 
$$

$$
= -\sum_{k=1}^{\kappa}\frac{Q_k'}{x-Q_k} - \frac{1}{\kappa}\sum_{k=1}^{\kappa}(x-Q_k) + J_0(t) - \phi'(t)/\phi + \sum_{k=1}^{\kappa}\frac{Q_k'}{x-Q_k},   
$$

\ni and using eq.~(3.28) leads to

$$
B_t \equiv B_1 + B_2 = -x + \frac{1}{\kappa}\sum_{k=1}^{\kappa}Q_k + J_0(t) - \phi'(t)/\phi = -x + \frac{1}{\kappa}\left(\frac{t^2}{2} + U(t)\right) - \phi'(t)/\phi.   \eqno(4.4)   
$$

\ni We are still to choose the other arbitrary function. In the relation following from formulas (2.29) and (3.14) (with $N=\kappa$),

$$
2b_+L_1 - 2B_1 = 2b_1 = \frac{1}{\kappa}\sum_{k=1}^{\kappa}\frac{\kappa Q_k' + t - Q_k^2 - 2R_k}{x-Q_k} - \frac{2}{\kappa}\sum_{k=1}^{\kappa}Q_k - J_0(t)   \eqno(4.5)   
$$

\ni (recall that $R_k = \sum_{j\neq k}1/(Q_k-Q_j)$, see eq.~(3.15)), $L_1$ and $B_1$ can be chosen in many different ways but a special choice below seems the best. Since we can separate in $L_1, L_2, B_1, B_2$ the trace parts determined by eqs.~(4.3), (4.4) as

$$
L_1 = \frac{-v + L_d}{2} = \frac{x^2-t + L_d}{2}, \ \ \ \ \ L_2 = \frac{-v - L_d}{2} = \frac{x^2-t - L_d}{2},
$$

$$
B_1 = \frac{B_t + B_d}{2}, \ \ \ \ \ B_2 = \frac{B_t - B_d}{2},
$$

\ni eq.~(4.5) can be rewritten as an equation on the functions $L_d$ and $B_d$ (one of which is still arbitrary):

$$
b_+L_d - B_d = \frac{1}{\kappa}\sum_{k=1}^{\kappa}\frac{\kappa Q_k' - 2R_k}{x-Q_k} - \phi'(t)/\phi.   \eqno(4.6)   
$$

\ni From eq.~(4.1) we get

$$
\prt_xL_+ = \phi(t)\sum_{k=1}^{\kappa}\prod_{j\neq k}^{\kappa}(x-Q_j), \ \ \ \ \ \prt_{xx}L_+ = \phi(t)\sum_{k=1}^{\kappa}2R_k\prod_{j\neq k}^{\kappa}(x-Q_j), 
$$

$$
\kappa\prt_tL_+ = \kappa\phi'/\phi - \phi(t)\sum_{k=1}^{\kappa}\kappa Q_k'\prod_{j\neq k}^{\kappa}(x-Q_j),   \eqno(4.7)    
$$

\ni therefore eq.~(4.6) can also be rewritten in terms of $L_+$, the chosen polynomial in $x$:

$$
\prt_xL_+\cdot L_d + L_+\cdot\kappa B_d = \kappa\prt_tL_+ + \prt_{xx}L_+.   \eqno(4.8)   
$$

\ni One can choose the remaining arbitrary function $L_d$ in such a way as to cancel out the poles on the right-hand side of eq.~(4.6), i.e.

$$
L_d = -\frac{L_+}{\phi(t)}\cdot\sum_{k=1}^{\kappa}\frac{\kappa Q_k' - 2R_k}{(x-Q_k)\prod_{j\neq k}^{\kappa}(Q_k-Q_j)} = -\sum_{k=1}^{\kappa}(\kappa Q_k' - 2R_k)\prod_{j\neq k}^{\kappa}\frac{x-Q_j}{Q_k-Q_j}.   \eqno(4.9)   
$$



\ni Then $B_d$ is determined by eq.~(4.6):

$$
\kappa(B_d-\phi'(t)/\phi) = -\sum_{k=1}^{\kappa}\frac{\kappa Q_k' - 2R_k}{x-Q_k} + \prt_xL_+\cdot\sum_{k=1}^{\kappa}\frac{\kappa Q_k' - 2R_k}{\prt_xL_+(Q_k)(x-Q_k)},
$$

\ni where $\prt_xL_+(Q_k) \equiv \prt_xL_+(t, x=Q_k(t)) = \phi(t)\prod_{j\neq k}^{\kappa}(Q_k-Q_j)$, see the first formula in eq.~(4.7). Thus, one can see that the chosen $L_d$ is a polynomial in $x$ of degree $\kappa-1$, and $B_d$ is a polynomial of degree $\kappa-2$ since it is a rational function whose only apparent poles $\{Q_k\}$ cancel between the two sums and the polynomial $\prt_xL_+$ has degree $\kappa-1$. A more explicit expression for $B_d$ is

$$
\kappa B_d = \kappa \phi'(t)/\phi + \sum_{k=1}^{\kappa}\frac{\kappa Q_k' - 2R_k}{x-Q_k}\left(\sum_{l=1}^{\kappa}\frac{\prod_{j\neq l}^{\kappa}(x-Q_j)}{\prod_{j\neq k}^{\kappa}(Q_k-Q_j)} - 1\right).   \eqno(4.10)    
$$

\par The remaining entries of the Lax matrices are then given by

$$
L_- = \frac{1}{L_+}\left(\kappa b_1 - (\kappa b_+ + v)L_1 - L_1^2 - \prt_xL_1\right) = \frac{\prt_xv+v^2/2-\kappa B_t - \kappa B_d - \prt_xL_d - L_d^2/2}{2L_+},
$$

\ni i.e.

$$
L_- = -\frac{\kappa B_d + \prt_xL_d + L_d^2/2 + V(v)}{2L_+},   \eqno(4.11)    
$$

\ni where we denoted the explicitly $v$-dependent part as

$$
V(v) = \kappa B_t - \prt_xv - v^2/2 = -\left(\frac{x^4}{2} - tx^2 + (\kappa-2)x - U(t) + \frac{\kappa \phi'(t)}{\phi}\right),  \eqno(4.12)   
$$

\ni and 

$$
\kappa B_- = b_+L_- + \frac{\prt_xB_1 - \prt_tL_1}{L_+} = -\frac{2\prt_xL_+L_- + \kappa \prt_tL_d - \kappa \prt_xB_d}{2L_+}.   \eqno(4.13)  
$$

{\it Remark.} It turns out that the above $L_-$ and $B_-$ are also polynomial (for $\kappa \in \N$), which is not obvious from eqs.~(4.11) and (4.13) together with eqs.~(4.1), (4.9) and (4.10). We defer a detailed consideration of this to a future publication.
\par Finally recall that the matrices of the Lax pair consist of the calculated entries:

$$
L = \left(\begin{array}{cc} L_1 & L_+ \\ L_- & L_2 \end{array}\right) = \left(\begin{array}{cc} \frac{1}{2}(x^2-t+L_d) & \phi(t)\prod_{k=1}^{\kappa}(x-Q_k(t)) \\ -\frac{1}{2L_+}(\kappa B_d + \prt_xL_d + L_d^2/2 + V(v)) & \frac{1}{2}(x^2-t-L_d) \end{array}\right),   
$$

$$
B = \left(\begin{array}{cc} B_1 & B_+ \\ B_- & B_2 \end{array}\right) = 
$$

$$
= \left(\begin{array}{cc} \frac{1}{2}(-x+(U(t)+t^2/2)/\kappa-\phi'/\phi + B_d) & -\frac{\phi(t)}{\kappa}\sum_{k=1}^{\kappa}\prod_{j\neq k}^{\kappa}(x-Q_j) \\ -\frac{1}{2L_+}(2L_-\prt_xL_+ + \kappa \prt_tL_d - \kappa \prt_xB_d) & \frac{1}{2}(-x+(U(t)+t^2/2)/\kappa-\phi'/\phi - B_d)  \end{array}\right),
$$

\ni where functions $L_d(t)$, $B_d(t)$ and $V(v)$ are given by eqs.~(4.9), (4.10) and (4.12) above. The functions $Q_k$, $k=1,\dots,\kappa$, in the matrix entries satisfy eqs.~(3.29), (3.30), and function $U(t)$ satisfies eq.~(3.27). Then the first component $\F_\kappa$ of the eigenvector solution to the linear system (2.2) with $L$ and $B$ matrices above, solves the Fokker-Planck equation (3.0), which we call Quantum Painlev\'e II.

\section{Open problems}

We showed how to construct Lax pairs whose eigenvector components are probability distributions for beta-ensembles, for even integer $\beta$. This should be possible to generalize to the other values of $\beta$, achieving the classical integrability for all beta-ensembles and related models. 
\par Exact relation of the system of ODEs for functions $Q_k$ of section 3 to Painlev\'e II and its solutions remains to be worked out for $\kappa\equiv\beta/2 > 2$. 
\par Our approach can be applied to the other quantum Painlev\'e equations. Construction of Lax pairs for the special values of $\beta$, $\beta=2, 4$, i.e.~$\kappa=1, 2$,  gives the corresponding governing functions $b_+$ and $b_1$, which should suggest the ansatz for the other integer values of $\kappa=\beta/2$. 
\par The remarkable duality, i.e.~the symmetry of many formulas in $\beta$-ensemble theory w.r.t.~the change $\kappa \leftrightarrow 1/\kappa$ implies some relations of the results for integer $\kappa$ with those for $\kappa = 1/k, k\in \mathbb{N}$. Indeed, the quantum Painlev\'e equations for beta-ensembles introduced by Nagoya in~\cite{Nag11} give $1/\kappa$ in place of our $\kappa$, so our results here are in fact relevant for $\kappa = 1/k, k\in \mathbb{N}$ in application to the ensembles of~\cite{Nag11}, where $t$ has different meaning -- it is {\it coupling constant} in an eigenvalue measure potential, while here it is the (rescaled) spectral endpoint. This remarkable connection has to be better understood in all aspects, e.g.~in conjunction with the duality formulas of~\cite{Der09}.
\par Relations of $\tau$-functions of Painlev\'e equations with CFT recently established for $\beta=2$~\cite{GaIoLi} should be possible to extend to other values of $\beta$.
\par An important open problem for future work is to put our results in the more general framework of classical and quantum integrability relations, which is currently being built~\cite{ALTZ}, i.e.~to determine the exact places of $\beta$-related functions and equations in classical integrable hierarchies. The works on one-dimensional Toda hierarchy and generating functions similar to $\F_{\kappa}$ for $\kappa=1$~\cite{NaKaMaP2, NaKaMaP4, KMO07} should be also helpful in this respect. The last works together with our current result hint at a possibility to extend the Toda connection to other values of $\beta$, a question raised e.g.~in~\cite{MMShRes}. We defer this to a future work.
\par Another framework being developed recently is that of Macdonald processes~\cite{BC-Mac, BCGSh}, which can be considered e.g.~as a generalization of 1-parameter beta-ensemble problems to the 2-parameter problems related to Macdonald operators and functions. The appearance of Fredholm determinants there is in our opinion an indication of hidden classical integrability. Thus, another big open problem is to generalize our considerations here to the context of Macdonald stochastic particle processes, where again quantum integrability should be equivalent to classical. The connection of the two last mentioned frameworks should tie it all together and provide a more general quantum-classical correspondence of integrable systems.

\bigskip

{\bf\large Acknowledgments.} I would like to thank B. Rider, R. Maier and W. Hereman for interesting and useful discussions, and the referee for careful reading and multiple suggestions to improve the text and correct some errors. The financial support by NSF grant DMS-0645756 is gratefully acknowledged. Also I would like to thank MSRI at Berkeley, CA, for support during my stay there in Fall 2010 there as a postdoctoral fellow in Special Program on Random Matrices and Random Processes, where my interest in the current problem setting was initiated.

\bigskip

\section*{Appendix A: proof of compatibility of equations (3.22), (3.23) and (3.24)}

Let us start with eq.~(3.22) in the form (using eq.~(3.28))

$$
\frac{t^2}{2} + U + \sum_{k=1}^{\kappa}Q_k = A_k(t-Q_k^2-A_k/2) - \sum_{j\neq k}^{\kappa}\frac{A_k-A_j}{Q_k-Q_j}.    \eqno(A1)
$$

\ni Differentiating eq.~(A1) and using eqs.~(3.21), 

$$
A_k = \kappa Q_k' + t - Q_k^2 - 2R_k,   \eqno(A2)   
$$

\ni and (3.15), we obtain

$$
\frac{1}{\kappa}\sum_{k=1}^{\kappa}A_k = (t-Q_k^2-A_k)A_k' + A_k\left(1-\frac{2}{\kappa}Q_k(A_k-t+Q_k^2+2R_k)\right) - \sum_{j\neq k}^{\kappa}\frac{A_k'-A_j'}{Q_k-Q_j} + 
$$

$$
+ \frac{1}{\kappa}\sum_{j\neq k}^{\kappa}\frac{(A_k-A_j)(A_k-A_j + Q_k^2 - Q_j^2 + 2(R_k-R_j))}{(Q_k-Q_j)^2}.   \eqno(A3)
$$

\ni Introducing the quantities 

$$
2X_k = \kappa A_k' + 2Q_kA_k,   \eqno(A4)
$$

\ni we rewrite eq.~(A3) as

$$
\sum_{k=1}^{\kappa}A_k = 2(t-Q_k^2-A_k-R_k)X_k + \sum_{j\neq k}^{\kappa}\frac{2X_j}{Q_k-Q_j} + \kappa A_k + 
$$

$$
+ \sum_{j\neq k}^{\kappa}\left(\frac{(A_k-A_j)(A_k-A_j + Q_k^2 - Q_j^2 + 2(R_k-R_j))}{(Q_k-Q_j)^2} - \frac{2(Q_jA_j+Q_kA_k)}{Q_k-Q_j}\right).   \eqno(A5)
$$

\ni Next we rearrange the terms as follows:

$$
\sum_{j\neq k}^{\kappa}\left(\frac{(A_k-A_j)(Q_k^2 - Q_j^2)}{(Q_k-Q_j)^2} - \frac{2(Q_jA_j+Q_kA_k)}{Q_k-Q_j}\right) = \sum_{j\neq k}^{\kappa}\frac{2A_j(Q_j^2 - Q_k^2)}{(Q_k-Q_j)^2} - \sum_{j\neq k}^{\kappa}(A_k-A_j),   \eqno(A6)
$$

\ni and use that

$$
\sum_{k=1}^{\kappa}A_k = \kappa A_k - \sum_{j\neq k}^{\kappa}(A_k-A_j),   \eqno(A7)
$$

\ni to reduce eq.~(A5) to

$$
0 = 2(t-Q_k^2-A_k-R_k)X_k + \sum_{j\neq k}^{\kappa}\frac{2X_j}{Q_k-Q_j} + 
$$

$$
+ \sum_{j\neq k}^{\kappa}\frac{(A_k-A_j)^2 + 2A_j(Q_j^2 - Q_k^2) + 2(A_k-A_j)(R_k-R_j)}{(Q_k-Q_j)^2}.  \eqno(A8)
$$

\ni Now we rearrange terms in the last sum as

$$
(A_k-A_j)^2 + 2A_j(Q_j^2 - Q_k^2) + 2(A_k-A_j)R_k = (A_j^2 + 2(Q_j^2-t)A_j) - (A_k^2 + 2(Q_k^2-t)A_k) + 
$$

$$
+ 2(A_k-A_j)(A_k+Q_k^2-t+R_k),   \eqno(A9)
$$

\ni and use eq.~(A1) implying that

$$
(A_j^2 + 2(Q_j^2-t)A_j) - (A_k^2 + 2(Q_k^2-t)A_k) = \sum_{l\neq k}^{\kappa}\frac{2(A_k-A_l)}{Q_k-Q_l} - \sum_{l\neq j}^{\kappa}\frac{2(A_j-A_l)}{Q_j-Q_l} \equiv 2(\Phi_k - \Phi_j),   \eqno(A10)
$$

\ni to finally rewrite eq.~(A8) as

$$
0 = 2(t-Q_k^2-A_k-R_k)\left(X_k - \sum_{j\neq k}^{\kappa}\frac{A_k-A_j}{(Q_k-Q_j)^2}\right) + \sum_{j\neq k}^{\kappa}\frac{2}{Q_k-Q_j}\left(X_j - \sum_{l\neq j}^{\kappa}\frac{A_j-A_l}{(Q_j-Q_l)^2}\right) + 
$$

$$
+ \sum_{j\neq k}^{\kappa}\frac{2}{Q_k-Q_j}\left(\frac{\Phi_k - \Phi_j - (A_k-A_j)R_j}{Q_k-Q_j} + \sum_{l\neq j}^{\kappa}\frac{A_j-A_l}{(Q_j-Q_l)^2}\right).   \eqno(A11)
$$

\ni To finish the proof, we will show that the last sum in eq.~(A11) is zero. Then eq.~(A11) is satisfied by putting

$$
X_k = \sum_{j\neq k}^{\kappa}\frac{A_k-A_j}{(Q_k-Q_j)^2},   \eqno(A12)
$$

\ni which is exactly eq.~(3.23) after recalling the definition eq.~(A2). To this end, we find

$$
\Phi_j + (A_k-A_j)R_j = \sum_{l\neq j}^{\kappa}\frac{A_j-A_l}{Q_j-Q_l} + \sum_{l\neq j}^{\kappa}\frac{A_k-A_j}{Q_j-Q_l} = \sum_{l\neq j}^{\kappa}\frac{A_k-A_l}{Q_j-Q_l},   \eqno(A13)
$$

\ni and the last sum in eq.~(A11) becomes

$$
\sum_{j\neq k}^{\kappa}\frac{2}{Q_k-Q_j}\left(\sum_{l\neq k}^{\kappa}\frac{A_k-A_l}{(Q_k-Q_j)(Q_k-Q_l)} - \sum_{l\neq k, j}^{\kappa}\frac{A_k-A_l}{(Q_k-Q_j)(Q_j-Q_l)} + \sum_{l\neq j}^{\kappa}\frac{A_j-A_l}{(Q_j-Q_l)^2}\right) = 
$$

$$
= \sum_{j\neq k}^{\kappa}\frac{2}{Q_k-Q_j}\sum_{l\neq k, j}^{\kappa}\left(\frac{A_k-A_l}{(Q_k-Q_j)(Q_k-Q_l)} - \frac{A_k-A_l}{(Q_k-Q_j)(Q_j-Q_l)} + \frac{A_j-A_l}{(Q_j-Q_l)^2}\right) = 
$$

$$
= \sum_{j\neq k}^{\kappa}\frac{2}{Q_k-Q_j}\sum_{l\neq k, j}^{\kappa}\left(\frac{A_j-A_l}{(Q_j-Q_l)^2} - \frac{A_k-A_l}{(Q_k-Q_l)(Q_j-Q_l)}\right) = 
$$

$$
= 2\sum_{j\neq k}^{\kappa}\sum_{l\neq k, j}^{\kappa}\frac{A_j(Q_k-Q_l) - A_l(Q_k-Q_j)}{Q_k-Q_j} = 0,
$$

\ni the last two equalities being true by antisymmetry of the summand w.r.t.~permutation of summation indices $j$ and $l$. This ends the proof. 

\section*{Appendix B: derivation of equations (3.29) and (3.30)}

We substitute $A_k$ from eq.~(3.21) into eq.~(3.23) to express it explicitly in terms of $Q_k$ variables only. After using

$$
\sum_{j\neq k}\frac{Q_k^2-Q_j^2}{(Q_k-Q_j)^2} = \sum_{j\neq k}\frac{Q_k+Q_j}{Q_k-Q_j} = 2Q_kR_k - \sum_{j\neq k}1 = 2Q_kR_k - (\kappa-1),
$$

\ni and after some cancellations eq.~(3.23) becomes

$$
\kappa^2 Q_k'' = -2Q_k(t-Q_k^2) + \kappa-2 - 4\sum_{j\neq k}\frac{R_k-R_j}{(Q_k-Q_j)^2}.
$$

\ni Since

$$
R_k-R_j = \sum_{l\neq k}^{\kappa}\frac{1}{Q_k - Q_l} - \sum_{l\neq j}^{\kappa}\frac{1}{Q_j - Q_l} = \frac{2}{Q_k-Q_j} + \sum_{l\neq k,j}^{\kappa}\left(\frac{1}{Q_k - Q_l} - \frac{1}{Q_j - Q_l}\right)
$$

\ni and

$$
\sum_{j\neq k}^{\kappa}\frac{1}{(Q_k - Q_j)^2}\sum_{l\neq k,j}^{\kappa}\left(\frac{1}{Q_k - Q_l} - \frac{1}{Q_j - Q_l}\right) = -\sum_{j\neq k}^\kappa\sum_{l\neq k,j}^{\kappa}\frac{1}{(Q_k-Q_j)(Q_k - Q_l)(Q_j - Q_l)} = 0
$$


\ni (the last equality above follows from the anti-symmetry of the summand in the indices $j$ and $l$), one obtains eq.~(3.29).   
\par Then we will rewrite eq.~(3.22) more explicitly. After substituting eq.~(3.28), eq.~(3.22) can be rewritten as

$$
\frac{(\kappa Q_k' - t + Q_k^2)A_k}{2} - \sum_{j\neq k}^{\kappa}\frac{A_j}{Q_k - Q_j} + \sum_{k=1}^{\kappa}Q_k + \frac{t^2}{2} + U = 0.   \eqno(B1)   
$$

\ni Substituting eq.~(3.21) into eq.~(B1) gives

$$
\frac{(\kappa Q_k')^2}{2} - \frac{(t - Q_k^2)^2}{2} - \sum_{j\neq k}^{\kappa}\frac{\kappa Q_k'-t+Q_k^2}{Q_k - Q_j} - \sum_{j\neq k}^{\kappa}\frac{\kappa Q_j'+t-Q_j^2-2R_j}{Q_k - Q_j} + \sum_{k=1}^{\kappa}Q_k + \frac{t^2}{2} + U = 0,  \eqno(B2)   
$$

\ni or

$$
\frac{(\kappa Q_k')^2}{2} + tQ_k^2 - \frac{Q_k^4}{2} - \sum_{j\neq k}^{\kappa}\frac{Q_k^2-Q_j^2}{Q_k - Q_j} + \sum_{j=1}^{\kappa}Q_j - \sum_{j\neq k}^{\kappa}\frac{\kappa Q_k' + \kappa Q_j'}{Q_k - Q_j} + 
$$

$$
+ 2\sum_{j\neq k}^{\kappa}\frac{1}{Q_k - Q_j}\sum_{l\neq j}^{\kappa}\frac{1}{Q_j - Q_l} + U = 0.  \eqno(B3)  
$$

\ni Since

$$
- \sum_{j\neq k}^{\kappa}\frac{Q_k^2-Q_j^2}{Q_k - Q_j} + \sum_{j=1}^{\kappa}Q_j = - \sum_{j\neq k}^{\kappa}(Q_k+Q_j) + \sum_{k=1}^{\kappa}Q_k = -(\kappa-1)Q_k + Q_k = -(\kappa-2)Q_k,   \eqno(B4)  
$$

\ni and

$$
\sum_{j\neq k}^{\kappa}\frac{1}{Q_k - Q_j}\sum_{l\neq j}^{\kappa}\frac{1}{Q_j - Q_l} = -\sum_{j\neq k}^{\kappa}\frac{1}{(Q_k - Q_j)^2} + \sum_{j\neq k}^{\kappa}\sum_{l\neq j,k}^{\kappa}\frac{1}{(Q_k-Q_j)(Q_j - Q_l)},   \eqno(B5)   
$$

\ni eq.~(B3) (equivalent to eq.~(3.22)) finally becomes the eq.~(3.30).




\end{document}